# Anion charge-lattice volume dependent Li ion migration in compounds with the face-centered cubic anion frameworks

Zhenming Xu[a], Xin Chen[b], Ronghan Chen[a], Xin Li[c], Hong Zhu[a,*]

a. *University of Michigan–Shanghai Jiao Tong University Joint Institute, Shanghai Jiao Tong University, 800, Dongchuan Road, Shanghai, 200240, China*

b. *School of Materials Science and Engineering, Shanghai Jiao Tong University, 800, Dongchuan Road, Shanghai, 200240, China*

c. *John. A. Paulson School of Engineering and Applied Sciences, Harvard University, 29 Oxford St, Cambridge, Massachusetts, 02138, USA*

**Abstract:** The proper design principles are essential for the efficient development of superionic conductors. However, the existing design principles are mainly proposed from the perspective of crystal structures. In this work, the face-centered cubic (*fcc*) anion frameworks were creatively constructed to study the effects of anion charge and lattice volume on the stability of lithium ion occupation and lithium ion migration. Both the large negative anion charges and large lattice volumes would increase the relative stabilities of lithium-anion tetrahedron, and make Li ions prefer to occupy the tetrahedral sites. For a tetrahedral Li ion migration to its adjacent tetrahedral site through an octahedral transition state, the smaller the negative anion charge is, the lower the lithium ion migration barrier will be. While for an octahedral Li ion migration to its adjacent octahedral site through a tetrahedral transition state, the larger negative anion charge is, the lower the lithium ion migration barrier will be. New design principles for developing superionic conductors with the *fcc* anion framework were proposed. Low Li ion migration barriers would be achieved by adjusting the non-lithium elements within the same crystal structure framework to obtain the desired electronegativity difference between the anion element and non-lithium cation element.

[*]Corresponding author:
E-mail: hong.zhu@sjtu.edu.cn

## 1. Introduction

Safety is the most important concern when using the commercial lithium ion batteries (LIBs) in the application scenarios of the large-scale energy storage, such as electric vehicles. Replacing the currently employed flammable liquid electrolytes in LIBs with the solid-state electrolyte (SSE) materials and collocating with the Li metal anodes to construct the all-solid-state lithium ion batteries (ASSLIBs) not only could solve the battery safety issues, but also remarkably enhance the energy density of battery systems[1]. Correspondingly, the construction of practical ASSLIBs needs SSE materials to achieve Li ion fast conduction with low activation energies (less than 300 meV) and a high Li ionic conductivity ($10^{-3}$-$10^{-2}$ S $cm^{-1}$) at room temperature. So far, some superionic conductors, such as $Li_7La_3Zr_2O_{12}$[2], $Li_{1+x}Al_xTi_{2-x}(PO_4)_3$[3] oxides and $Li_{10}GeP_2S_{12}$[4], $Li_7P_3S_{11}$[5] sulfides have been widely studied as the SSE materials, and the state-of-the-art ionic conductivities of 12-17 mS $cm^{-1}$ at room temperature are experimentally realized in $Li_{10}GeP_2S_{12}$ and $Li_7P_3S_{11}$ sulfides.

To efficiently develop more advanced superionic conductors for ASSLIBs, the better understanding of fast ion migration mechanism in the state-of-the-art superionic conductors and the development of proper design principles are quite essential. Ceder et. al. have proposed an important design principle for superionic conductors that the body-centred cubic (*bcc*) anion framework with face-sharing lithium-anion tetrahedra allows the low activation energy of Li ion migration[6], which is successfully guiding the high-throughput screening of new superionic conductors[7]. G. Hautier et. al. found the distorted lithium-sulfur polyhedrons in $LiTi_2(PS_4)_3$ provide the smooth energy landscape combining small activation barriers with numerous migration paths, and proposed the design concept of “frustrated energy landscape” for superionic conductors[8]. The mobile species with unfavored coordination environments are correlated with high ionic conductivities[9], which is consistent with the concept of “frustrated energy landscape”. Furthermore, the high-throughput screening of fast lithium ion conductors by Xiao et al. shows the activation energies of Li ion migration in the olivine-structures are lower than those of the layered- and even spinel-structures[10].

However, the existing design principles of the face-sharing lithium-anion tetrahedron and frustrated energy landscape are mainly proposed from the perspective of crystal structure without the considerations of other factors, e.g. binding strength between the migrating Li ion and its adjacent anions, and the polarizability of anion[11]. It is generally accepted that the Columbic force dominates the interaction between Li cation and its adjacent anion framework in ionic materials[12]. In the spinel $LiMn_2O_4$, the different valence states of Mn ions and their arrangements surrounding Li ions have important effects on the activation barrier of Li migrations[12a]. Our previous study of the chalcopyrite-structured $LiMS_2$ (M are transition metals, from Ti to Ni) materials with the same crystal structure demonstrates that the larger negative anion charges resulted from the larger electronegativity difference between M and S elements would increase the activation barrier for Li ion migration between the two adjacent tetrahedral (*Tet*) sites through an octahedral (*Oct*) transition state, namely *Tet-Oct-Tet* pathway[12b]. In addition, Mo et al. constructed an artificial face-centered cubic (*fcc*) anion sublattice of the monovalent $S^-$ in comparison with the bivalent $S^{2-}$ with a constant lattice volume, and found activation barrier for Li-ion migration along the *Tet-Oct-Tet* pathway in the monovalent $S^-$ sublattice is smaller than that of the bivalent $S^{2-}$ sublattice[12c]. On the contrary, in $Li_3MI_6$ (M=Sc, Y and La) compounds with stable octahedral Li occupations, the larger negative I anion charges would lower the activation barrier for Li ion migration along the *Oct-Tet-Oct* pathways[13]. The questions then become, why anion charge shows the reverse influence on Li migration barrier for the *Tet-Oct-Tet* and *Oct-Tet-Oct* pathway? Are there any connections among Li occupation pattern, anion charge and lattice volume? Therefore, in this work, we made efforts to further understand the roles of anion charge as well as lattice volume on the Li ion occupations and Li ion migrations, and proposed new design principles for developing superionic conductors.

## 2. Computational details

This work is based on the density functional theory (DFT) calculations performed by using the Vienna *ab*-initio Simulation Package (VASP) software. The interaction between ion cores and valence electrons described by the projector augmented wave (PAW) method[14]. The generalized gradient approximation (GGA)[15] in the form of Perdew–Burke–Ernzerhof (PBE) exchange functional[16] was used to solve the quantum states of electron. The plane-wave energy cutoff is set to 500 eV. The Monkhorst–Pack method[17] with $1\times1\times2$ *k*-point mesh is employed for the Brillouin zone sampling of the super lattice. The convergence criterions of energy and force are set to $10^{-5}$ eV/atom and 0.01 eV/Å, respectively. The anion charges of lithium compounds were calculated by using the Atoms in Molecules method (Bader charge analysis)[18]. The energy variations and migration barriers of lithium ion migration in the *fcc*-type anion frameworks with 48 anions (Figure S4) are calculated by the nudged elastic band (NEB) method[19]. The anion charges are changed by the uniform background charge of the framework system. Only the one migrating Li ion is allowed to relax, while the other anions are fixed in their initial positions, and this method can be also found in Ceder's work[6].

## 3. Results and discussion

The topologies of the close-packed anions of the common lithium ionic conductor materials can be approximately classified into the *fcc*, body-centered cubic (*bcc*) and hexagonal close-packed (*hcp*) frameworks[6]. The anion frameworks of $LiCoO_2$, $Li_2MnO_3$, $Li_4Ti_5O_{12}$, $Li_2S$, $LiTiS_2$ and $Li_3YBr_6$[20] can be exactly matched to the *fcc* types. For $Li_7P_3S_{11}$ and $Li_{10}GeP_2S_{12}$, the S anion sublattice can be roughly mapped to *bcc* lattices with some distortions. In both $\gamma$-$Li_3PS_4$ and $Li_4GeS_4$, the S anion sublattices can be closely matched to *hcp* arrays[6]. In the aforementioned lithium compounds as well as more than half lithium compounds in the Materials Project (MP) database, Li ions mainly occupy the tetrahedral or octahedral sites, forming the stable tetrahedral or octahedral lithium-anion polyhedrons, as shown in the distribution of lithium coordination environments (Figure S1 in Supporting Information). We find that there are pairs of the adjacent anion tetrahedron and octahedron sharing a triangular face in the *fcc*, *bcc* and *hcp* anion frameworks. Li ion migration between two adjacent *Tet* and *Oct* sites can be regarded as the half migration path for the *Tet-Oct-Tet* or *Oct-Tet-Oct* hoppings. Considering the *fcc* and *hcp* anion frameworks are much more common than *bcc* anion arrangement (Figure S2), in this work, by the DFT calculations, we mainly focused on the *fcc* anion framework to efficiently investigate the Li occupation patterns as well as Li migration between two adjacent *Oct* and *Tet* sites (two face-sharing octahedron and tetrahedron) from a new perspective of the effect of anion charge as well as lattice volume, from which the new design principles for efficiently searching superionic conductors were proposed. The anion charge and lattice volume dependent Li occupation pattern and Li migration in the *hcp* anion arrangements will be further studied in another work.

### 3.1 Anion charge and lattice volume dependent Li occupation and migration

First, we have calculated the anion Bader charges and lattice volumes of some stable lithium oxides and sulfides from the MP database, to determine the reasonable value ranges of anion charge and lattice volume, as listed in Table S1 and S2 in Supporting Information. Figure S3 shows that the scatter distributions of anion charge and lattice volume of these lithium oxides and sulfides around the fitted straight lines, approximately demonstrating a positive correlation between anion charge and lattice volume. For the convenience of making good comparison, the lattice volumes are averaged to each anion from the volumes of unit cell. Then, an artificial *fcc*-type anion framework with 48 anions and one single Li ion (Figure S4) was built to simulate the Li ion migration between two adjacent *Oct* and *Tet* central sites, as the local structure shown in Figure 1a. This computational strategy can make us directly capture the effect of anion charge and lattice volume, which has been successfully used by Ceder et al.[6] Then, the nudged elastic band (NEB) calculations were performed to monitor the energy variations for Li ion migration from an *Oct* site to its adjacent *Tet* site with respect to different anion charges and lattice volumes, as schematically shown in Figure 1b. Here, different kinds of anion were considered, including O, S, F, Cl, Br and I anions, and the calculated results of Li ion migration barriers ($E_m$) and the energy differences ($E_{tet-oct}$) between the *Tet* Li site and *Oct* Li site are shown as the heat maps in Figure 2, S3 and S4. Anion charge and lattice volume have significant impacts on the $E_m$ and $E_{tet-oct}$ values for both the chalcogen (Figure 2) and halogen (Figure S5 and S6) anion systems. In addition, a consistent mechanisms of anion charge and lattice volume on $E_m$ and $E_{tet-oct}$ are observed for different anion systems. The variation trends of $E_m$ and $E_{tet-oct}$ with respect to different anion charge and lattice volume are much more interesting than their absolute values. Taking O anion framework as an example (Figure 2a), within different O lattice volume regions, the O anion charges have different effects on $E_m$ values for Li ion migration. However, for a specific O lattice volume, the increasing negative O anion charges consistently reduce $E_{tet-oct}$ values (Figure 2c), and stabilizing the $LiO_4$ tetrahedron. At a constant O anion charge, $E_{tet-oct}$ values vary from positive to negative, and the relative stabilities of $LiO_4$ tetrahedron are gradually increased when the O lattice volumes get larger. In addition, the S, F, Cl, Br and I anion frameworks show the similar effects of anion charge and lattice volume on Li ion migration and the relative stabilities of lithium-anion tetrahedron. In summary, both the larger negative anion charges and large lattice volumes make Li ions prefer to occupy *Tet* sites.

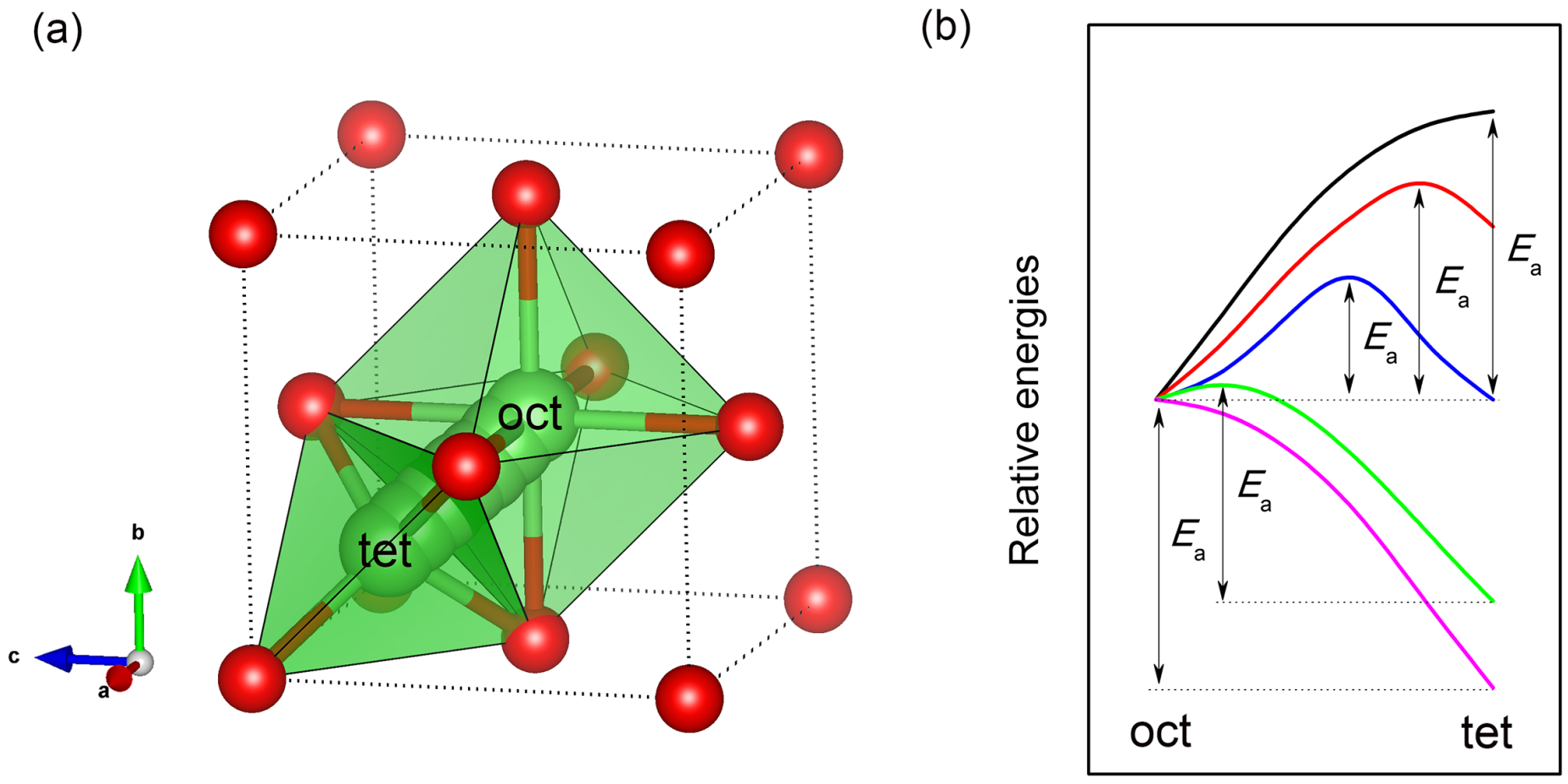

Figure 1. (a) Li ion migration between the two adjacent *Oct* and *Tet* sites in an artificial *fcc*-type anion framework, which is the local structure from the anion framework in Figure S4. (b) schematic diagram of the energy variations of Li ion migration between the two adjacent *Oct* and *Tet* central sites with respect to different anion charges and lattice volumes. The Li ions and anions are colored red and green, respectively.

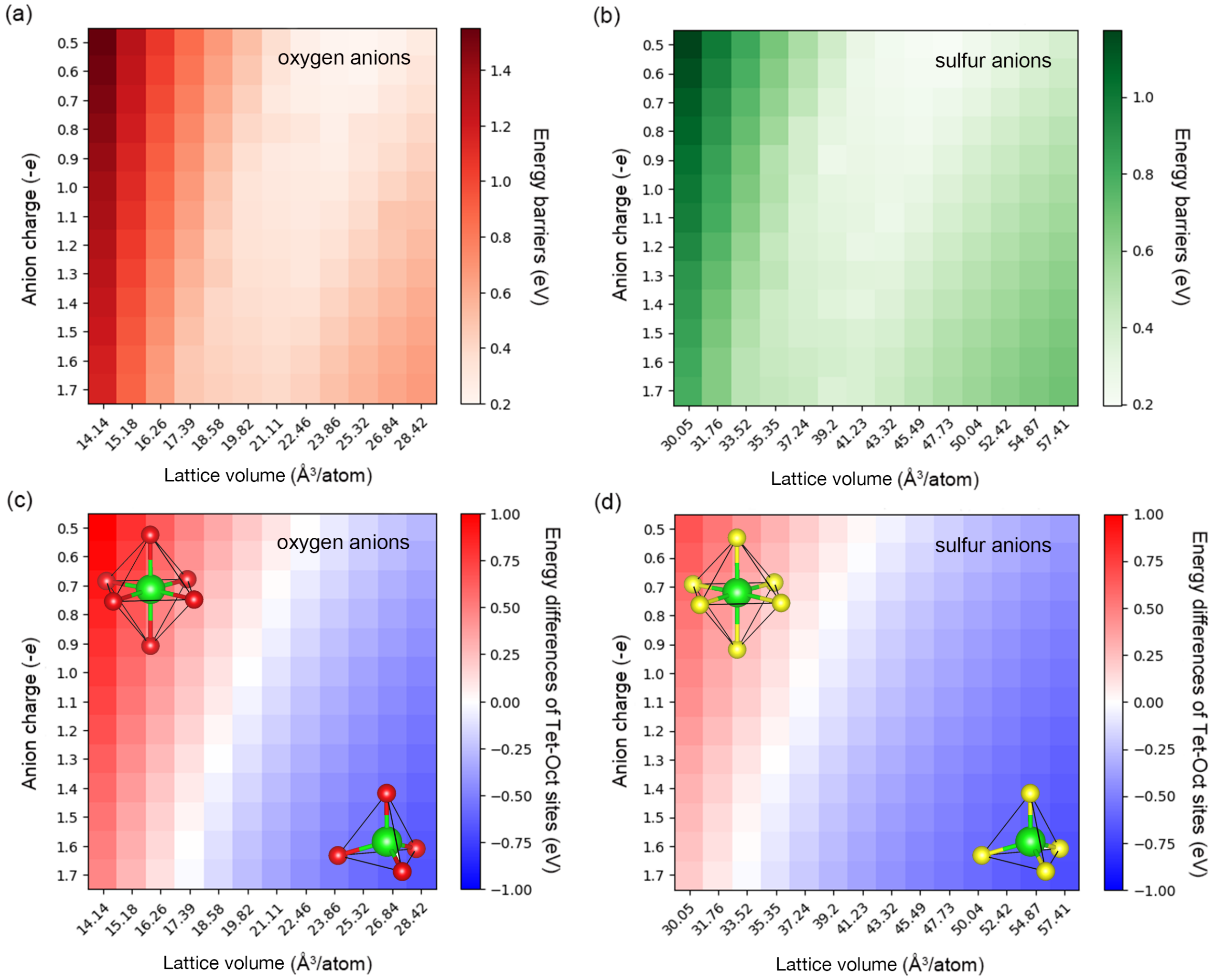

Figure 2. Heat maps of the calculated $E_m$ of Li ion migration between the two adjacent *Oct* and *Tet* central sites in the artificial *fcc*-type (a) oxygen and (b) sulfur anion frameworks, and the energy differences between the *Tet* Li site and *Oct* Li site in *fcc*-type (c) oxygen and (d) sulfur anion frameworks with respect to different anion charges and lattice volumes, respectively.

For the conveniences of clear insights into the effects of anion charge, we chose three representative lattice volumes of the O anion systems from the heat maps in Figure 2, and the O anion charge dependent energy variations of Li ion migration between the two adjacent *Oct* and *Tet* central sites and the corresponding $E_m$ at three fixed lattice volumes are shown in Figure 3. When the O lattice volume is small, e.g. with a value of ~16.26 Å$^3$/atom (16.41 Å$^3$/atom for R$\bar{3}$m-$LiCoO_2$, Table S1), the relative energies of $LiO_4$ are all higher than those of $LiO_6$ ($E_{tet\text{-}oct} > 0$, Figure 3a), indicating Li ions are most stable in the *Oct* sites, which are consistent with the fact that lithium oxides with smaller O lattice volumes (16-18 Å$^3$/atom) show the octahedral Li occupations (Table S1 and Figure S3a). In addition, the larger negative O anion charges would lower the relative energies of $LiO_4$, and hence reduce the corresponding $E_m$ for Li ion migration in these O anion frameworks with smaller O lattice volumes (Figure 3d). For the O anion systems with medium lattice volumes, e.g. ~19.82 Å$^3$/atom (19.86, 20,42 and 20.43 Å$^3$/atom for $Li_2SiO_3$, $Li_2FeSiO_4$ and $Li_2MnSiO_4$, respectively, Table S1), with the negative O anion charges increasing from -0.5 to -1.7*e*, the $E_m$ values would first decrease and then increase (Figure 3d). This is because the relative energies of $LiO_4$ are higher than those of $LiO_6$ ($E_{tet-oct} > 0$, Figure 3b) for the systems with the smaller negative O anion charges ($q_O < -1.0$). While the larger negative O anion charges ($q_O > -1.0$) make the relative energies of $LiO_4$ lower than those of $LiO_6$ ($E_{tet-oct} < 0$) and the *Oct* sites no longer stable. At a larger lattice volume, e.g. ~23.86 Å$^3$/atom (23.67 and 24.97 Å$^3$/atom for $Li_5AlO_4$ and $Li_2O$, respectively, Table S1), the relative energies of $LiO_4$ are lower than those of $LiO_6$ ($E_{tet-oct} < 0$, Figure 3c), indicating Li ions prefer the *Tet* sites at the large lattice volumes, as shown in Table S1 and Figure S3a that the lithium oxides with larger O lattice volumes (> 21 Å$^3$/atom) showing the tetrahedral Li occupations. Moreover, we found that Li ion migrations in these O anion frameworks with large O lattice volumes (~23.86 Å$^3$/atom) become more difficult with the increase of negative O anion charges (Figure 3d). Viewed from Figure 3d that when the negative O anion charges are less than -1.2*e*, the increased O lattice volumes would reduce $E_m$ for Li ion migration. While the increasing O lattice volumes make $E_m$ first decrease and then increase when the negative O anion charges are more than -1.2*e*, which is consistent with the earlier study on the *fcc* $S^{2-}$ sublattice by Ceder et al[6]. In summary, the larger negative anion charges would deliver a high $E_m$ for the tetrahedral Li ion migration along the *Tet*-*Oct*-*Tet* pathways, but a lower $E_m$ for the octahedral Li ion migration along the *Oct*-*Tet*-*Oct* pathways.

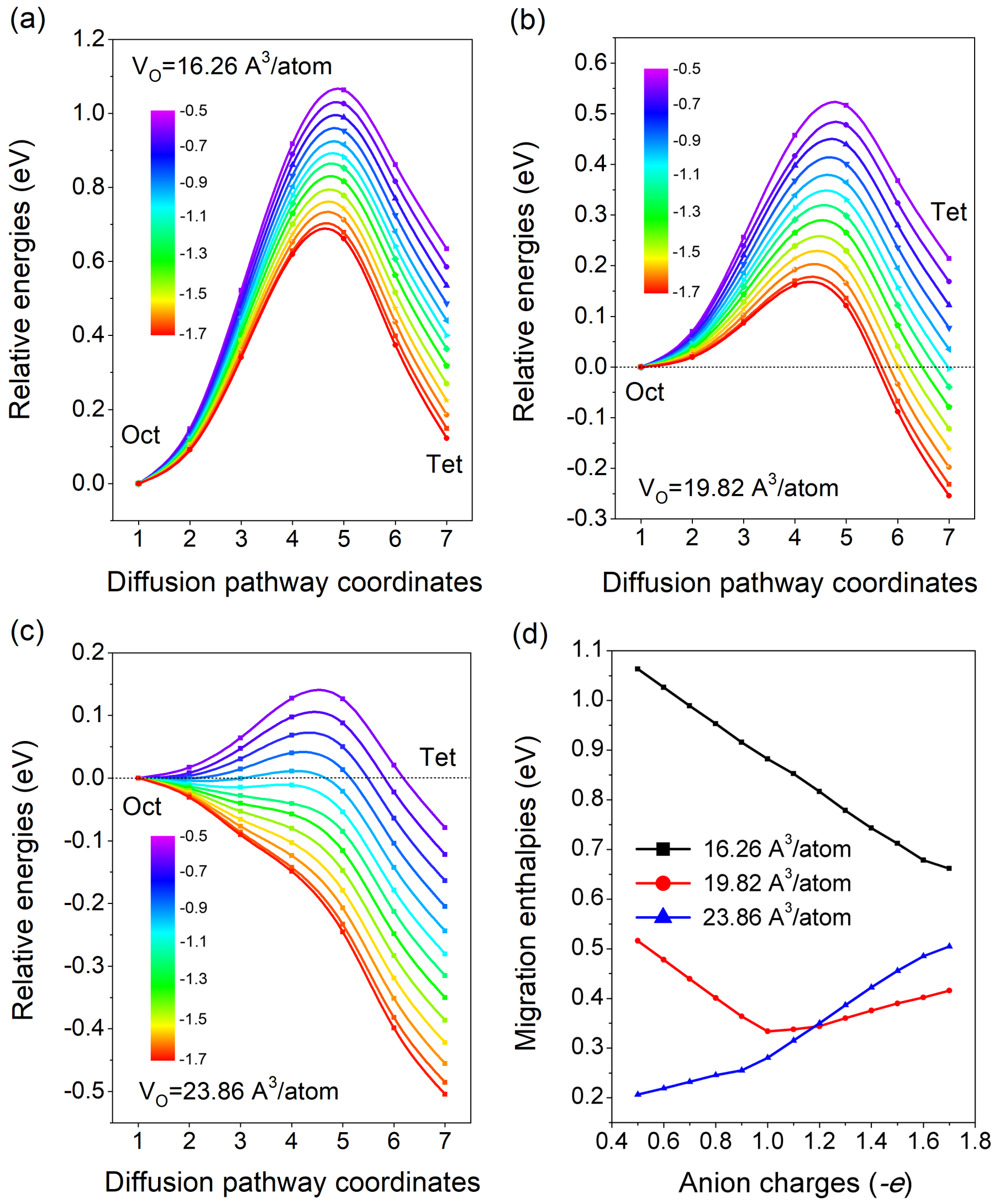

Figure 3. DFT calculations monitored energy variations and $E_m$ of Li ion migration between the two adjacent *Oct* and *Tet* central sites in the artificial *fcc*-type oxygen anion lattices with respect to different anion charges and constant lattice volumes. Energy variations of Li ion migration in the oxygen anion lattices with lattice volumes of (a) $V_o$= 16.26 Å$^3$/atom; (b) $V_o$= 19.82 Å$^3$/atom; (c) $V_o$= 23.86 Å$^3$/atom; (d) $E_m$ of Li-ion migration in oxygen anion framework.

## 3.2 Model validation

The energy barrier and jump distance for Li ion migration are determined by the total energy landscape of Li. The total energy landscape of Li ion in an ionic solid depends on the electrostatic interaction between Li ion and other ions, which can be further divided into a short-range Li-anion attractive interaction and a longer-range Li-cation repulsive interaction[8]. The Li-anion attractive interaction is modulated by the high-frequency alternations of the stable Li occupation sites separated by the energy barriers that Li needs to overcome when squeezing through a small bottleneck to reach the adjacent stable site. While the Li-cation repulsive interactions show much longer modulations on the order of the distance between two cations. The resulting total energy landscapes are mainly set by the Li-anion interactions, so Li ion migration in an ionic compound can be approximatively reduced to Li ion migration in an anion framework model. It is also noted that the Li-cation repulsive interactions also contribute to the total energy landscape to some extents, and the weight of the Li-cation interaction in setting the total energy landscape is set by the arrangements and valance states of cation.

Combining the anion framework model (heat maps in Figure 2) with the calculated Bader charges and lattice volumes (listed in Table S1 and S2) of some lithium oxides and sulfides, the corresponding $E_m$ were predicted, as shown in Figure 4 and Table S3, in comparison with some available NEB and *ab*-initio molecular dynamics (AIMD) calculated $E_m$ of lithium compounds with *fcc* anion frameworks. It is found that there are some discrepancies between the predicted $E_m$ and NEB calculated results for some lithium compounds. These deviations may come from the non-negligible Li-cation interactions and distorted anion frameworks of some real lithium compounds resulting in different total energy landscapes than those set by our orderly anion framework model. It's worth noting that most of these NEB calculated $E_m$ are path and migration mechanism dependent, which are not considered in our anion framework model. In addition, even for $LiCoO_2$, $LiMn_2O_4$ and $LiTiS_2$ with only a specific Li ion migration path, we find that different researchers get quite different NEB results (Table S3). Therefore, making comparison between our model predicted $E_m$ and the NEB calculated $E_m$ for a specific path is not meaningful. While the $E_m$ from AIMD simulations can be regarded as the statistical average for Li ion migration along different paths, and validating our model predicted results of lithium compounds by using the corresponding AIMD simulation calculated $E_m$ are meaningful. Unfortunately, the AIMD calculated results for lithium compounds are very rare, and only solid-state electrolytes of $\gamma$-$Li_3PO_4$ and $\beta$-$Li_3PS_3$ are available. It is delightful that our anion framework model predicted $E_m$ of 0.32 eV for $\beta$-$Li_3PS_3$ agrees well with the corresponding AIMD calculated results of 0.29 and 0.31 eV. On the other hand, the predicted $E_m$ of 0.50 eV for $Li_2S$ are much close to the NEB calculated results of 0.47 and 0.48 eV[6, 21], indicating that Li ion migration in $Li_2S$ is mainly dominated by the attractive interactions between $Li^+$ ion and its neighboring S anions, and the electrostatic repulsion interactions of $Li^+$-$Li^+$ don't affect Li ion migration too much. Because the $Li^+$-$Li^+$ repulsion interactions are much smaller than those of $Li^+$-M (M are cations with high valance state such $Mn^{4+}$, $P^{5+}$) in the ternary and quaternary lithium compounds. The good accuracy of the predicted $E_m$ for $Li_2S$ firmly validates our anion framework model again.

Additionally, it can be seen from Figure 4 that $E_m$ from both model prediction and NEB calculations of Li four-coordinated compounds are relatively smaller than those of Li six-coordinated compounds, as least for the above-mentioned oxides and sulfides, which is consistent with the fact of most superionic conductors showing Li tetrahedral occupations, such as $Li_3PO_4$, $Li_3PS_4$, $Li_7P_3S_{11}$ and $Li_{10}GeP_2S_{12}$. Most importantly, beyond the compounds in Table S3, $E_m$ of Li migration in other lithium compounds with face-sharing tetrahedron and octahedron can be predicted by our anion framework model, associated with known anion charges and lattice volumes. The AFLOW database contains many material compounds with structure and Bader charge information[22], therefore, it is feasible to screen lithium superionic conductors with *fcc* anion frameworks by combining our model with the AFLOW database without extra DFT calculations.

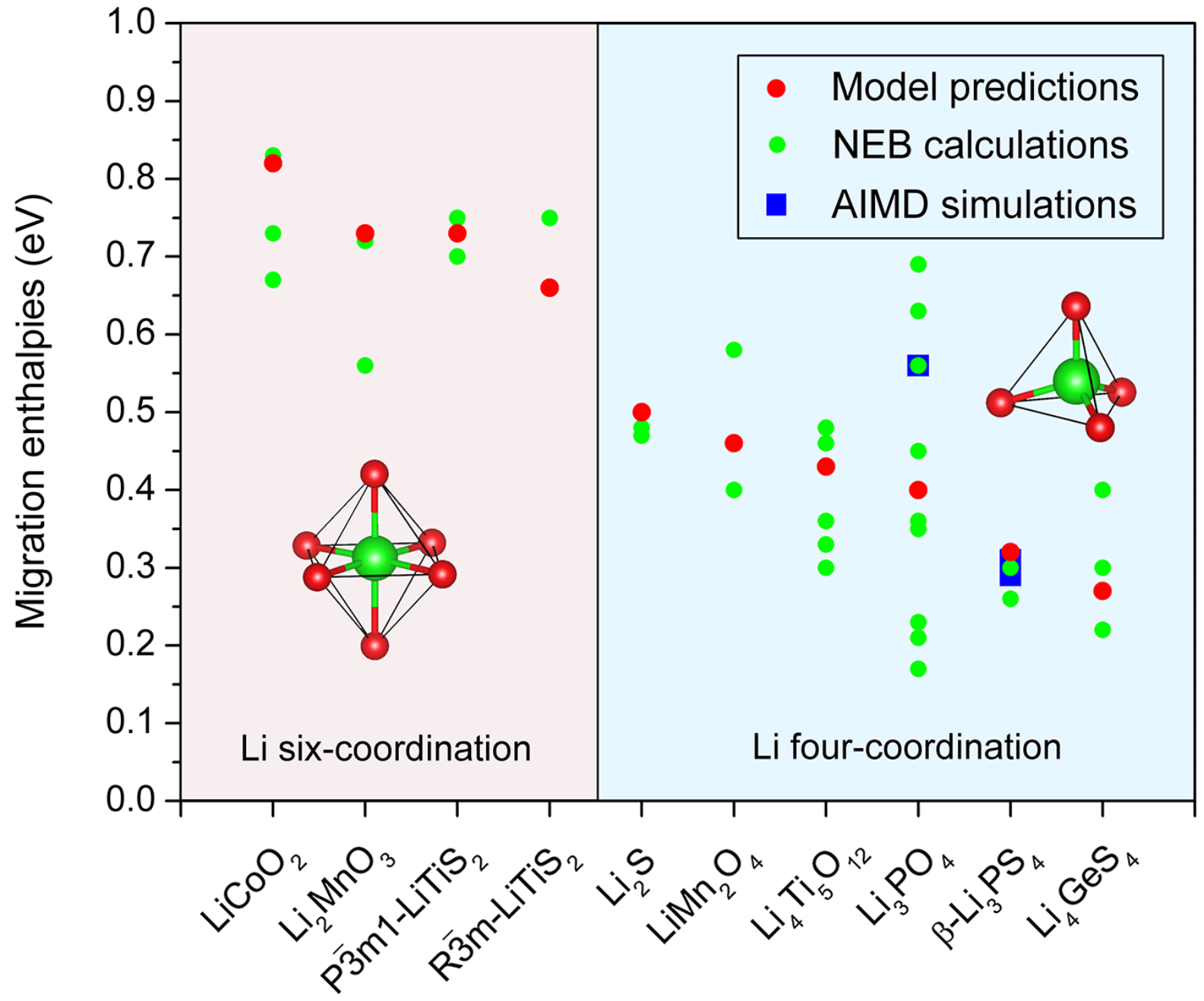

Figure 4. Comparisons among the predicted $E_m$ from the anion framework model, and NEB calculated $E_m$ for Li ion migration by the *Tet-Oct-Tet* or *Oct-Tet-Oct* pathways, and $E_m$ from *ab*-initio molecular dynamics simulations for some real lithium compounds with *fcc* anion frameworks. The corresponding $E_m$ data in this figure are also listed in Table S3.

The above anion framework model analyses about the change trends of $E_m$ with respect to anion charge and lattice volume are also confirmed by some reported materials. Our previous work on the chalcopyrite-structured $LiMS_2$ (M are transition metals, from Ti to Ni) materials with tetrahedral Li occupations shows that the smaller negative S anion charges resulted from the smaller electronegativity difference between transition metal and sulfur element would lead to lower $E_m$ for Li ion migration along the *Tet-Oct-Tet* pathways[12b]. Mo et al. found $E_m$ for the tetrahedral Li ion migration along the *Tet-Oct-Tet* pathways in a *fcc* monovalent $S^-$ anion framework is much lower that of the bivalent $S^{2-}$ anion framework with the same lattice volumes[12c]. For the spinel structured $LiAlCl_4$, $Li_2MgCl_4$ and $Li_2MgBr_4$ with tetrahedral Li occupations, the very active Mg and Al elements enable larger negative anion charges, eventually showing higher $E_m$ for the *Tet-Oct-Tet* Li ion migration. Similar effect can be also found in $Li_{10}MP_2S_{12}$ (M = Ge and Sn) materials with tetrahedral Li occupations. The higher electronegativity of Ge *vs.* Sn (2.0 *vs* 1.7[23]) give rise to less electron densities on S anions in $Li_{10}GeP_2S_{12}$, leading to the smaller negative anion charges, and thereby show relatively lower $E_m$ compared to $Li_{10}SnP_2S_{12}$[24], which are in good accordance with the AIMD simulations by S. P. Ong et al[25]. The above reported lithium compounds with tetrahedral Li occupations consistently obey the rule of the smaller negative anion charges leading to higher $E_m$ for the tetrahedral Li ion migration, proposed in the foregoing model analyses of Figure 3c. On the other hand, our previous research on the lithium iodides[13], $Li_3MI_6$ (M=Sc, Y and La) with octahedral Li occupations, shows that the largest I anion negative charges of $Li_3LaI_6$ resulted from the most active La (Pauling electronegativity $\chi_A$, Sc ($\chi_A = 1.36$) > Y ($\chi_A = 1.22$) > La ($\chi_A = 1.10$)[26]) lead to the lowest phonon DOS center of Li and smallest $E_m$ for Li ion migration along the *Oct-Tet-Oct* pathways, which are also in good agreement with the foregoing model analyses of Figure 3a. We also find $E_m$ change of Li ion migration along the *Oct-Tet-Oct* pathways in the gradually charged $Li_xCoO_2$[27], $P\bar{3}m1$-$Li_xTiS_2$[28] and $P6_3/mmc$ -$Na_xCoO_2$[29] cathodes match our anion charge-lattice volume map, that is with more Li or Na extraction from these layered structures, the lattice parameter *c* as well as the anion charge would decrease to some extents[28, 30], making the values of anion charge-lattice volume locate at the more top left portion in the anion charge-lattice volume heat maps (Figure 2) and eventually increasing $E_m$. In total, the anion charge-lattice volume maps (change trends of $E_m$ with respect to anion charge and lattice volume) of anion framework model are reasonable and creditable, although the predicted absolute values of $E_m$ from anion framework model may differ from the NEB data especially for those electrolytes with high Li-cation repulsive interactions or large anion framework distortions.

### 3.3 New principles for developing superionic conductors with *fcc* anion frameworks

In a ternary, quaternary and even more polynary alkali metal compounds, the anion charges are usually affected by the electronegativity of the non-alkali metal elements, as confirmed by some previous work[12b, 24]. The atomic radius and valence electron configuration of the non-alkali metal element determines its coordination environment and the crystal volume, eventually affecting the corresponding lattice volumes. The above anion framework model analyses clearly show that anion charge and lattice volume significantly affect alkali metal ion occupation and ion migration. It is expected to achieve low $E_m$ for alkali metal ion migration by adjusting the non-alkali metal element within the same crystal structure framework. Here, based on the above findings, general principles for developing new ternary ABC type lithium, sodium or even multivalent metal superionic conductors with *fcc* anion frameworks can be summarized: (i) for the superionic conductors with stable A ion octahedral occupation sites, the large electronegativity difference between the anion element C and six-coordinated non-mobile cation element B is essential for achieving excellently fast A ion migration, as shown in Figure 5a, and the corresponding non-mobile cation element B should give preference to the elements located at the left bottom of the periodic table with small electronegativity, as shown in Figure 5b. The chemical components of the recent reported two superionic conductors, $Li_3YCl_6$ and $Li_3YBr_6$ with Li octahedral occupations[20], are completely in conformity with this octahedron principle, and their lithium ionic conductivities can be further enhanced by doping with rare-earth metal elements, whose electronegativity values are less than Y element; (ii) for the superionic conductor with stable A ion tetrahedral occupation sites, the small electronegativity difference between the anion element C and four-coordinated non-mobile cation element B is essential for achieving excellently fast A ion migration, as shown in Figure 5a, and the corresponding non-mobile cation element B should give preference to the elements located at the right top of the periodic table of elements with large electronegativity, which are close to but less than that of C element, as shown in Figure 5b. The chemical components of the most superionic conductors with Li tetrahedral occupations, such as $Li_3PS_4$ and $Li_7P_3S_{11}$ with parts of *Tet-Oct-Tet* lithium migration pathways, perfectly fit with this tetrahedron principle. We hope that these two guiding principles will contribute to the design and optimization of superionic conductors.

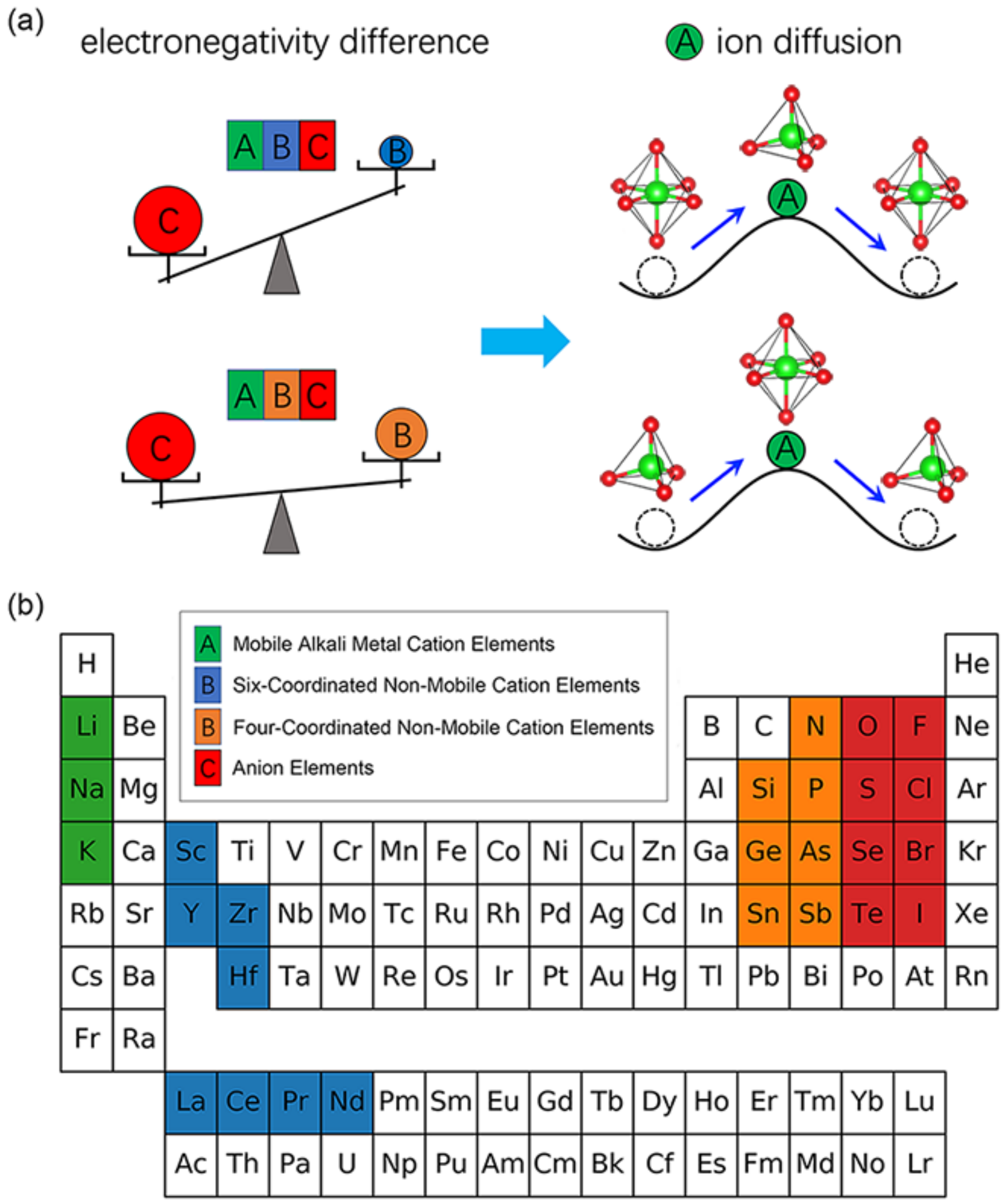

Figure 5. Design principles for fast A ion migration in the ABC ternary compounds, (a) schematic diagrams of the effects of the electronegativity differences between non-mobile cation elements B and anion elements C on A ion migration, (b) the recommended choices of the non-mobile cation element B in the periodic table of element for achieving fast A ion migration.

## 4. Conclusions

In summary, the *fcc* anion framework model show that anion charge and lattice volume significantly affect lithium ion occupation and ion migration, which is confirmed by many reported materials. Both the larger negative anion charges and large lattice volumes would enhance the relative stabilities of the tetrahedral Li occupation. For tetrahedral Li ion migration along the *Tet-Oct-Tet* pathways through an *Oct* transition state, the smaller negative anion charge is, the lower the lithium ion migration barrier is. While for octahedral Li ion migration along the *Oct-Tet-Oct* pathways through a *Tet* transition state, the larger negative anion charge is, the lower the lithium ion migration barrier is. Our anion framework model can be used for screening lithium superionic conductors with the *fcc* anion frameworks. Most importantly, new design principles for developing advanced superionic conductors with the *fcc* anion frameworks were proposed. Adjusting the non-mobile cation element within the same crystal structure framework to obtain the desired electronegativity differences between the anion element and non-mobile cation element, eventually achieving low $E_m$ for ion migration.

## Author's information

Corresponding Authors

*E-mail: hong.zhu@sjtu.edu.cn

Notes

The authors declare no competing financial interest.

## Acknowledgements

This work is supported by the National Natural Science Foundation of China (51602196), Shanghai Automotive Industry Corporation (1714), and Materials Genome Initiative Center at Shanghai Jiao Tong University. Z. Xu is much grateful to the support of the China Scholarship Council (CSC, scholarship No. 201906230117). All simulations were performed at the Shanghai Jiao Tong University High Performance Computing Center. We also thank Dr. S.H. Bo for the discussions.

# Supporting Information

## Anion charge-lattice volume dependent Li ion migration in compounds with the face-centered cubic anion frameworks

Zhenming Xu[a], Xin Chen[b], Ronghan Chen[a], Xin Li[c], Hong Zhu[a,*]

a. *University of Michigan–Shanghai Jiao Tong University Joint Institute, Shanghai Jiao Tong University, 800, Dongchuan Road, Shanghai, 200240, China*
b. *School of Materials Science and Engineering, Shanghai Jiao Tong University, 800, Dongchuan Road, Shanghai, 200240, China*
c. *John. A. Paulson School of Engineering and Applied Sciences, Harvard University, 29 Oxford St, Cambridge, Massachusetts, 02138, USA*

[*]Corresponding author:
E-mail: hong.zhu@sjtu.edu.cn

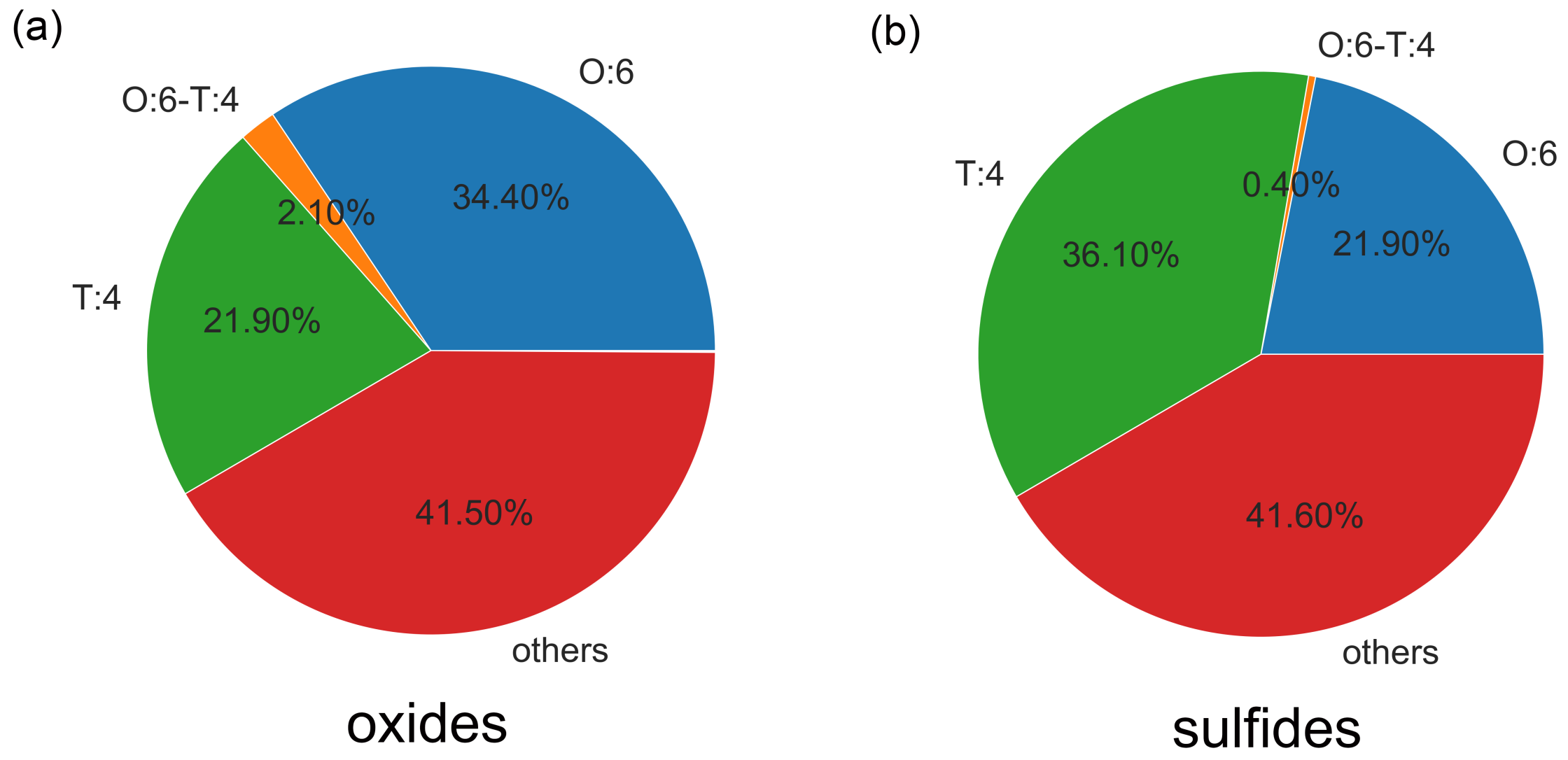

Figure S1. Distributions of lithium coordination environments for (a) oxides and (b) sulfides in the Materials Project (MP) database originating from the Inorganic Crystal Structure Database (ICSD). O:6 refers to a compound with only octahedral Li coordination, T:4 refers to a compound with only tetrahedral Li coordination, O:6-T:4 refers to a compound with both octahedral and tetrahedral Li coordination, and "others" refer to any other local environments in a compound, such as S:1 (single neighbor), L:2 (linear), A:2 (angular), TL:3 (trigonal plane), TY:3 (trigonal non-coplanar), S:4 (square plane), SS:4 (see-saw), S:5 (square-pyramidal), T:5 (trigonal bipyramid), T:6 (trigonal prism), PB:7 (pentagonal bipyramid), C:8 (cube), SA:8 (square antiprism), DDPN:8 (dodecahedron with triangular faces), HB:8 (hexagonal bipyramid) and C:12 (cuboctahedral). A structure with a part of O:6 or T:4 is also labeled as "other" type. The analyses of lithium coordination environment were performed by the ChemEnv code[1] as integrated in the Pymatgen package[2]. Only the stable binary, ternary and quaternary lithium compounds with energy above hull less than 0.05 eV/atom were considered, a total of 4334 lithium oxides and 235 lithium sulfides.

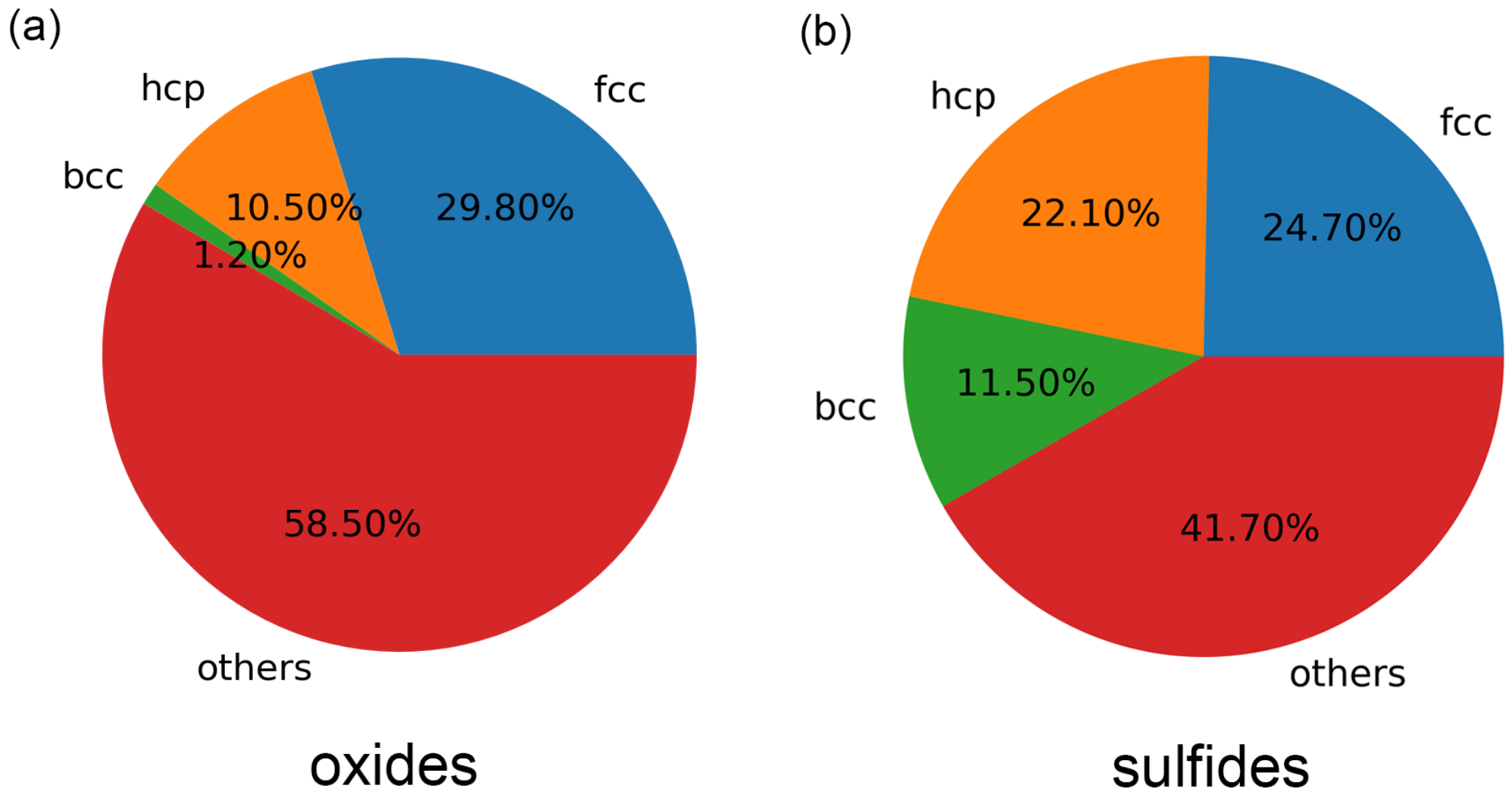

Figure S2. Distributions of the matched *fcc*, *bcc* and *hcp* anion frameworks for the slightly distorted (a) oxides and (b) sulfides in the MP database originating from ICSD. Others are the compounds with great anion distortions, exceeding the strict matching tolerances (fractional length tolerance = 0.1, site tolerance = 0.2, angle tolerance = 3 degrees), and may belong to the highly distorted *fcc*, *bcc* and *hcp* anion frameworks. The matchings of anion frameworks were performed by the StructureMatcher code[1] as integrated in the Pymatgen package[2]. Only the stable binary, ternary and quaternary lithium compounds with energy above hull less than 0.05 eV/atom were considered for matching, a total of 4334 lithium oxides and 235 lithium sulfides.

Table S1. Anion Bader charges ($q$), lattice volumes (V), and Li coordination number (CN) for some common lithium oxides from the MP database, and most of them show *fcc*-type oxygen anion frameworks.

| **Compounds** | **MP-ID** | $q$ ($e$) | **V** (Å$^3$/atom) | **CN** | **Compounds** | **MP-ID** | $q$ ($e$) | **V** (Å$^3$/atom) | **CN** |
|---|---|---|---|---|---|---|---|---|---|
| $LiMnO_2$ | mp-37620 | -1.266 | 16.276 | 6 | $Li_3PO_4$ | mp-13725 | -1.57 | 19.865 | 4 |
| $Li_2NiO_3$ | mp-566008 | -1.055 | 16.361 | 6 | $Li_2CO_3$ | mp-3054 | -1.322 | 20.118 | 4 |
| $LiFeO_2$ | mp-851027 | -1.090 | 16.385 | 6 | $Li_2SiO_3$ | mp-5012 | -1.634 | 20.208 | 4 |
| $LiCoO_2$ | mp-24850 | -1.064 | 16.411 | 6 | $LiSbO_3$ | mp-770932 | -1.19 | 20.26 | 4 |
| $Li_4Mn_5O_{12}$ | mp-691115 | -1.032 | 16.431 | 4, 6 | $Li_2HfO_3$ | mp-755352 | -1.414 | 20.396 | 6 |
| $Li_7Co_5O_{12}$ | mp-771536 | -1.064 | 16.432 | 6 | $Li_2FeSiO_4$ | mp-764790 | -1.535 | 20.421 | 4 |
| $LiFePO_4$ | mp-19017 | -1.475 | 16.693 | 6 | $Li_2MnSiO_4$ | mp-566680 | -1.561 | 20.43 | 4 |
| $Li_2MnO_3$ | mp-18988 | -1.212 | 16.72 | 6 | $LiPO_3$ | mp-29195 | -1.522 | 20.489 | - |
| $LiMnPO_4$ | mp-18997 | -1.499 | 17.05 | 6 | $LiV(PO_3)_4$ | mp-32492 | -1.461 | 20.683 | 4 |
| $LiNiO_2$ | mp-866271 | -1.084 | 17.129 | 6 | $Li_2SnO_3$ | mp-3540 | -1.361 | 20.757 | 6 |
| $LiNiO_2$ | mp-25592 | -1.068 | 17.459 | 6 | $Li_3BO_3$ | mp-27275 | -1.641 | 20.827 | 4, 5 |
| $LiTiO_2$ | mp-38280 | -1.371 | 17.723 | 6 | $Li_2Si_2O_5$ | mp-4117 | -1.619 | 21.098 | 4 |
| $Li_2FeO_3$ | mp-774155 | -1.115 | 17.881 | 6 | $Li_3BiO_4$ | mp-774702 | -1.206 | 21.423 | 6 |
| $Li_2TiO_3$ | mp-2931 | -1.307 | 18.208 | 6 | $LiAlO_2$ | mp-3427 | -1.655 | 21.633 | 4 |
| $LiCrO_2$ | mp-772487 | -1.242 | 18.349 | 6 | $Li_3CrO_4$ | mp-770632 | -1.145 | 21.913 | 4 |
| $LiV_2O_4$ | mp-19394 | -1.151 | 18.387 | 4 | $Li_8PtO_6$ | mp-8610 | -1.36 | 21.967 | 4, 6 |
| $Li_2RuO_3$ | mp-4630 | -1.131 | 18.509 | 6 | $Li_2GeO_3$ | mp-15349 | -1.346 | 22.196 | 4 |
| $Li_3RuO_4$ | mp-37692 | -1.13 | 18.544 | 6 | $Li_3VO_4$ | mp-19219 | -1.177 | 22.223 | 4 |
| $Li_4Ti_5O_{12}$ | mp-685194 | -1.187 | 18.588 | 4, 6 | $Li_4GeO_4$ | mp-4558 | -1.435 | 22.251 | 4 |
| $Li_2RhO_3$ | mp-754870 | -1.07 | 18.651 | 6 | $Li_2PbO_3$ | mp-22450 | -1.233 | 22.282 | 6 |
| $LiTi_2(PO_4)_3$ | mp-18640 | -1.39 | 18.669 | 6 | $LiGaO_2$ | mp-5854 | -1.332 | 22.476 | 4 |
| $LiMn_2O_4$ | mp-25015 | -1.075 | 18.672 | 4 | $Li_8SnO_6$ | mp-4527 | -1.527 | 22.781 | 4, 6 |
| $LiFeO_2$ | mp-851027 | -1.176 | 18.798 | 6 | $Li_8SiO_6$ | mp-28549 | -1.673 | 22.941 | 4 |
| $Li_2IrO_3$ | mp-532085 | -1.083 | 18.905 | 6 | $Li_2SeO_4$ | mp-4855 | -1.124 | 23.202 | 4 |
| $LiTi_2O_4$ | mp-5670 | -1.20 | 18.959 | 4 | $Li_4TiO_4$ | mp-9172 | -1.385 | 23.208 | 4 |
| $Li_3TaO_4$ | mp-3151 | -1.336 | 19.165 | 6 | $Li_2Ge_2O_5$ | mp-7998 | -1.294 | 23.602 | 4 |
| $Li_3NbO_4$ | mp-31488 | -1.295 | 19.182 | 6 | $Li_5AlO_4$ | mp-15960 | -1.682 | 23.893 | 4 |
| $LiCuO_2$ | mp-9158 | -1.028 | 19.537 | 6 | $Li_8CoO_6$ | mp-31531 | -1.549 | 24.278 | 4 |
| $LiRhO_2$ | mp-14115 | -1.031 | 19.656 | 6 | $Li_5BiO_5$ | mp-29365 | -1.34 | 24.356 | 5 |
| $Li_3SbO_4$ | mp-5769 | -1.325 | 19.725 | 4 | $Li_2O$ | mp-1960 | -1.683 | 24.969 | 4 |
| $LiNbO_2$ | mp-3924 | -1.252 | 19.806 | 6 | $Li_6CoO_4$ | mp-18925 | -1.538 | 25.533 | 4 |

Table S2. Anion Bader charges ($q$), lattice volumes (V), and Li coordination number (CN) for some common lithium sulfides from the MP database, and most of them show *fcc*-type sulfur anion frameworks.

| **Compounds** | **MP-ID** | $\boldsymbol{q}$ ($e$) | **V** (Å$^3$/atom) | **CN** | **Compounds** | **MP-ID** | $\boldsymbol{q}$ ($e$) | **V** (Å$^3$/atom) | **CN** |
|---|---|---|---|---|---|---|---|---|---|
| $LiCoS_2$ | mp-753946 | -0.747 | 28.619 | 6 | $LiYS_2$ | mp-15788 | -1.409 | 41.057 | 6 |
| $LiCrS_2$ | mp-4226 | -1.036 | 28.982 | 6 | $LiGaS_2$ | mp-3647 | -1.044 | 41.237 | 4 |
| $LiVS_2$ | mp-7543 | -1.116 | 29.612 | 6 | $Li_2TeS_3$ | mp-558731 | -0.979 | 41.275 | 6 |
| $Li_2(TaS_2)_3$ | mp-755664 | -1.059 | 31.627 | 6 | $Li_{10}Si(PS_6)_2$ | mp-720509 | -1.129 | 41.392 | 4, 6 |
| $P\bar{3}m1$-$LiTiS_2$ | mp-9615 | -1.221 | 31.889 | 6 | $\beta$-$Li_3PS_4$ | mp-985583 | -0.965 | 41.626 | 4 |
| $Li_5(NbS_2)_7$ | mp-767171 | -1.063 | 31.946 | 6 | $Li_{10}Ge(PS_6)_2$ | mp-696138 | -1.042 | 41.761 | 4, 6 |
| $Li_9(NbS_2)_{14}$ | mp-767218 | -1.021 | 32.017 | 6 | $LiGdS_2$ | mp-1222370 | -1.353 | 41.879 | 6 |
| $LiNbS_2$ | mp-7936 | -1.107 | 32.091 | 6 | $Li_{10}Sn(PS_6)_2$ | mp-721236 | -1.043 | 42.683 | 4, 6 |
| $R\bar{3}m$-$LiTiS_2$ | mp-1001784 | -1.236 | 32.2 | 6 | $Li_4GeS_4$ | mp-30249 | -1.192 | 42.831 | 4, 6 |
| $Li(NiS)_2$ | mp-769205 | -0.792 | 34.574 | 6 | $LiSbS_2$ | mp-1079885 | -1.040 | 42.894 | 6 |
| $Li_3NbS_4$ | mp-769032 | -1.124 | 34.682 | 6 | $Li_3BS_3$ | mp-5614 | -1.403 | 42.97 | 4 |
| $LiScS_2$ | mp-1001786 | -1.333 | 35.851 | 6 | $Li_4TiS_4$ | mp-766540 | -1.298 | 43.508 | 4 |
| $Li(ZrS_2)_2$ | mp-1222722 | -1.196 | 35.909 | 6 | $Li_3SbS_4$ | mp-756316 | -1.035 | 45.006 | 4 |
| $Li_2SnS_3$ | mp-1190364 | -1.119 | 37.132 | 6 | $Li_4SnS_4$ | mp-1195718 | -1.229 | 45.449 | 4 |
| $LiZnPS_4$ | mp-11175 | -0.755 | 38.497 | 4 | $Li_2S$ | mp-1153 | -1.717 | 46.821 | 4 |
| $Li_2US_3$ | mp-15885 | -1.241 | 39.374 | 6 | $LiInS_2$ | mp-1188392 | -1.045 | 47.024 | 4 |
| $LiErS_2$ | mp-15791 | -1.357 | 40.091 | 6 | $Li_4TiS_4$ | mp-861182 | -1.275 | 47.392 | 4 |
| $LiAlS_2$ | mp-1106183 | -1.541 | 40.49 | 4 | $Li_3CuS_2$ | mp-1177695 | -1.431 | 48.98 | 4 |

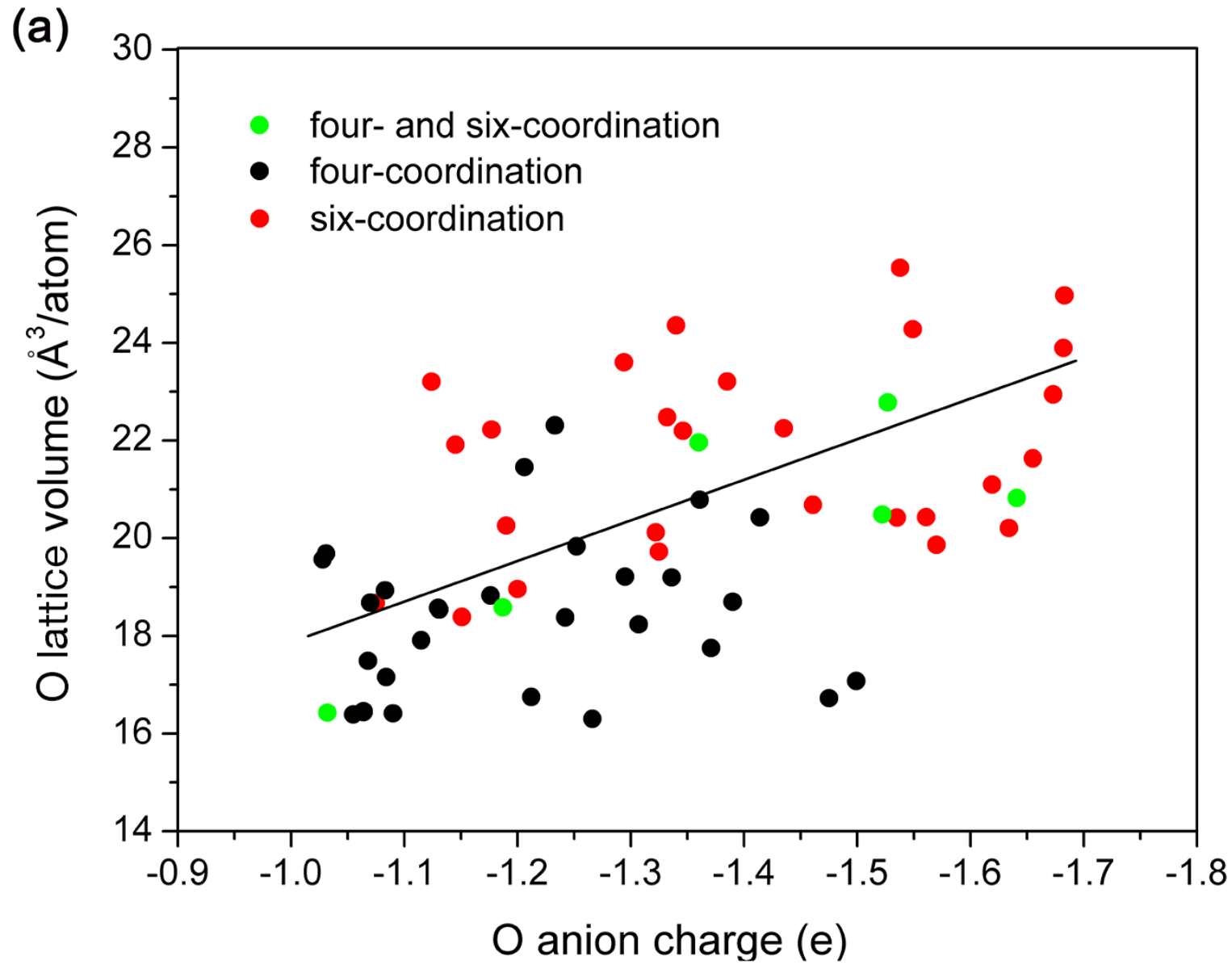

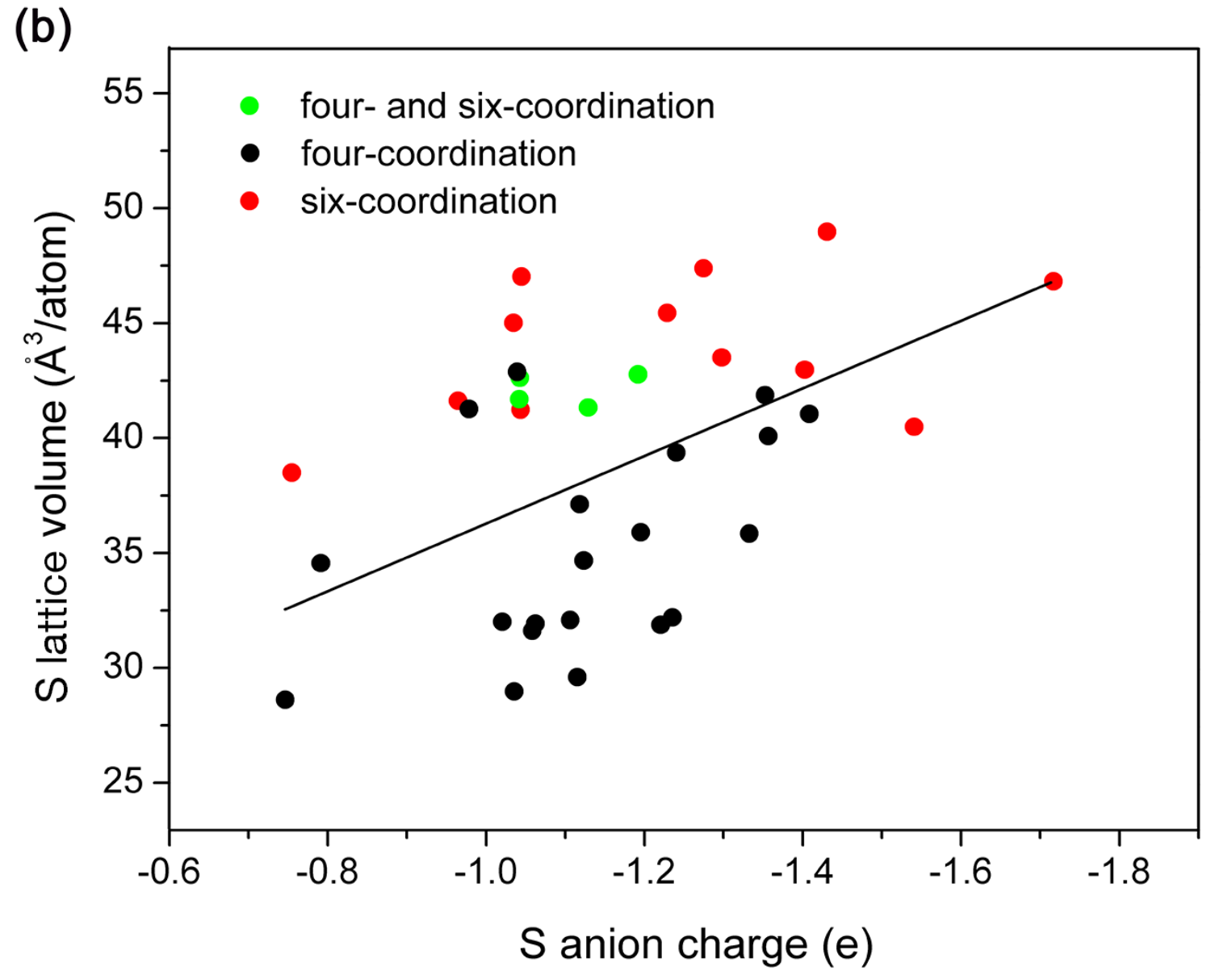

Figure S3. The scatter distributions of the anion charges and lattice volumes around the fitted straight lines for some common lithium (a) oxides and (b) sulfides listed in Table S1 and S2.

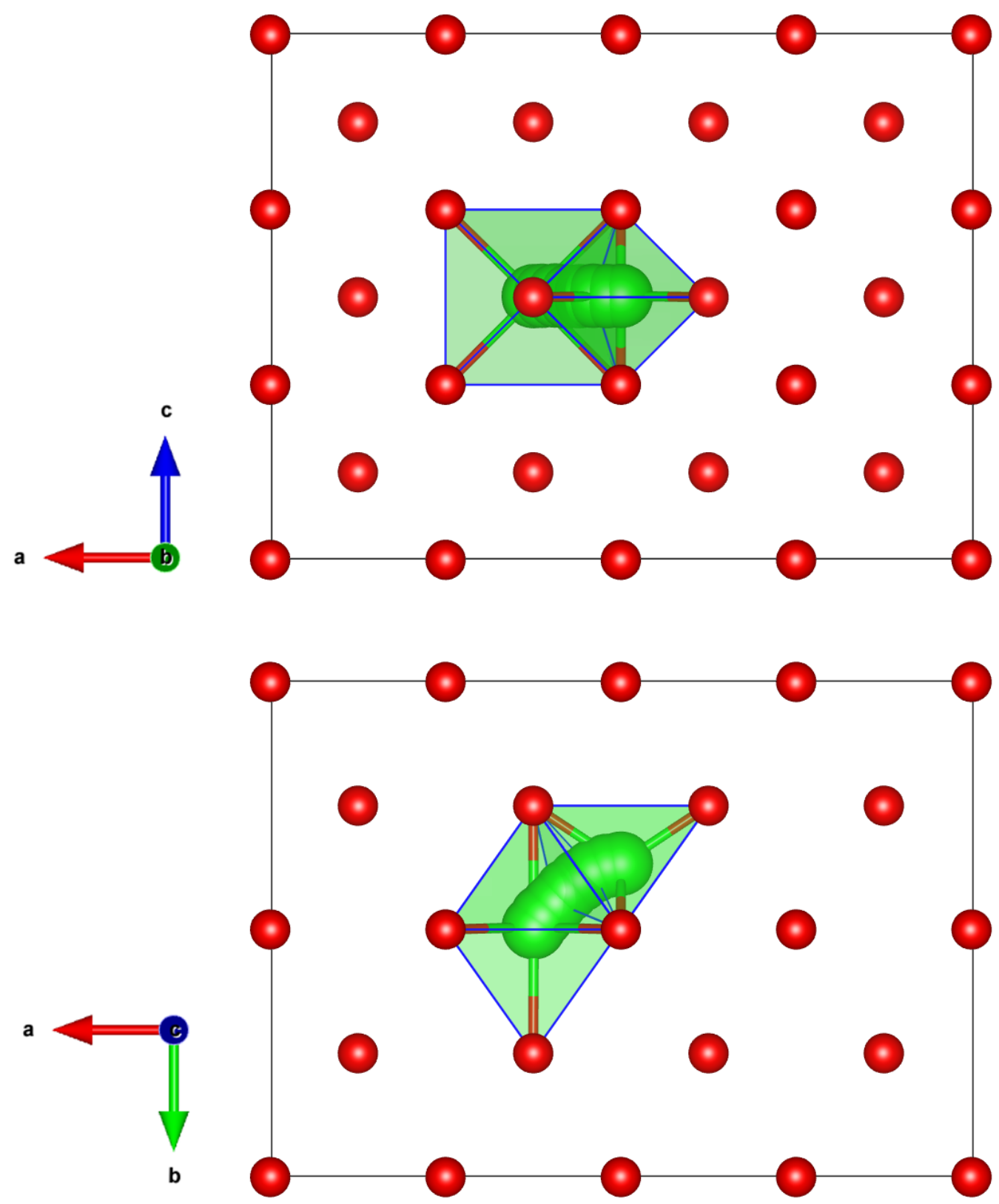

Figure S4. Structural model of Li ion diffusion from one octahedral site to its adjacent tetrahedral site with respect to different anion charges and lattice volumes in an artificial *fcc*-type anion sublattice with 48 anions. The anions are colored red, and the Li ions are colored green, respectively. Only the migrating Li ion is allowed to relax while the other anions are fixed to their initial positions.

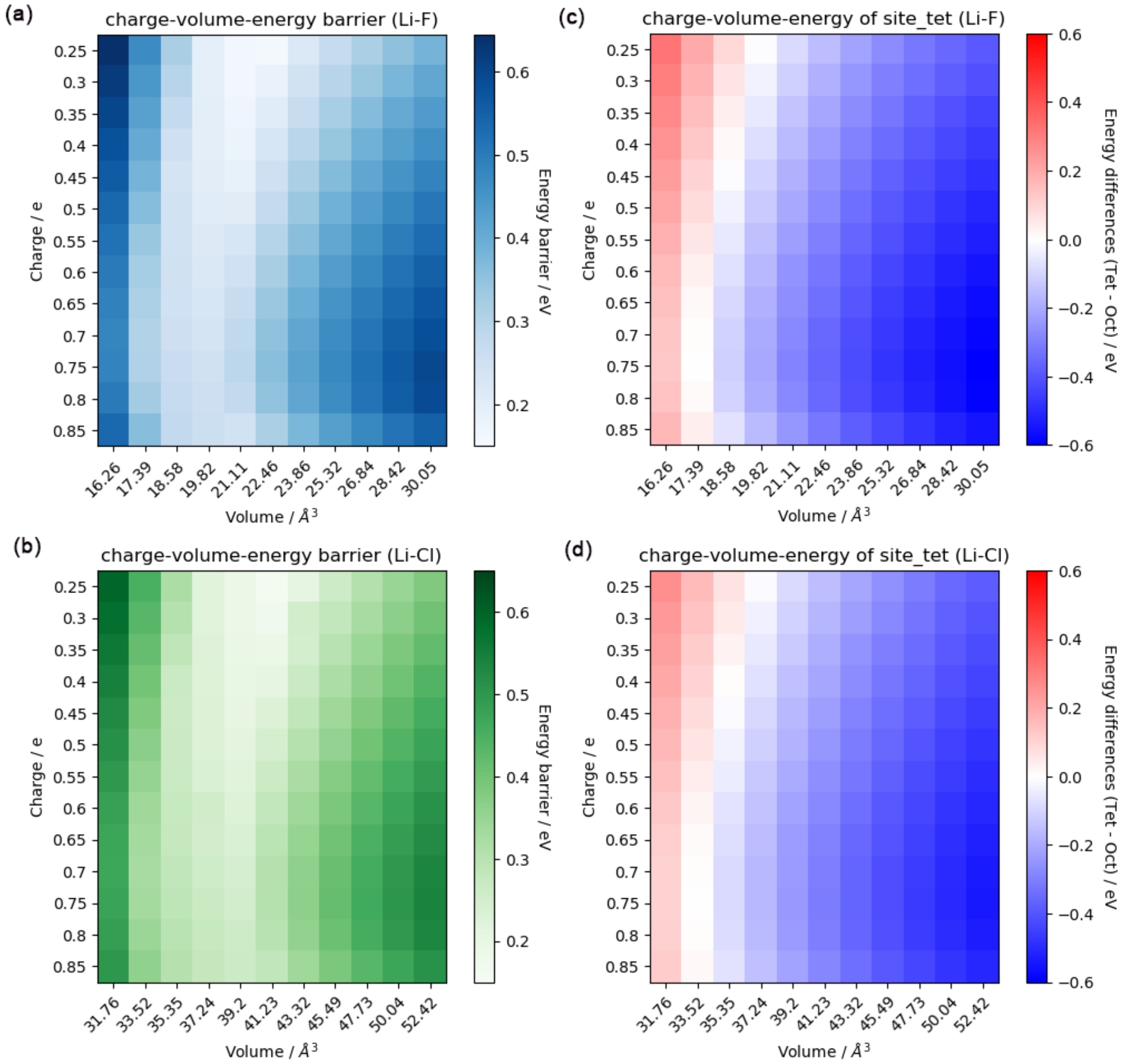

Figure S5. Heat maps of the calculated $E_m$ of Li ion migration between the two adjacent *Oct* and *Tet* central sites in the artificial *fcc*-type (a) fluorine and (b) chlorine anion lattices, and the energy differences between the *Tet* Li site and *Oct* Li site in an artificial *fcc*-type (c) fluorine and (d) chlorine anion lattice with respect to different anion charges and lattice volumes, respectively.

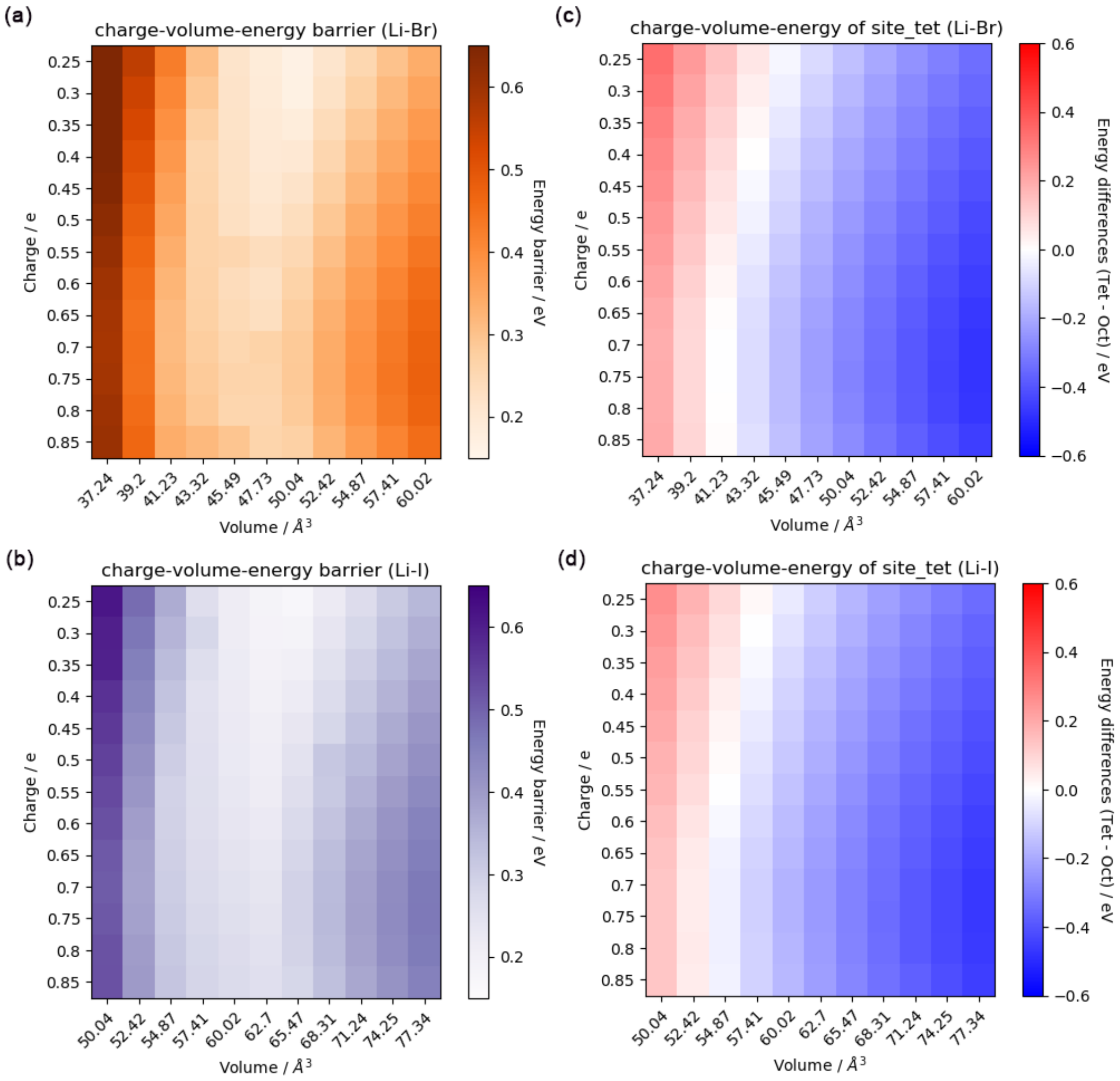

Figure S6. Heat maps of the calculated $E_m$ of Li ion migration between the two adjacent *Oct* and *Tet* central sites in the artificial *fcc*-type (a) bromine and (b) iodine anion lattices, and the energy differences between the *Tet* Li site and *Oct* Li site in an artificial *fcc*-type (c) bromine and (d) iodine anion lattice with respect to different anion charges and lattice volumes, respectively.

Table S3. Comparisons between the predicted $E_m$ (in eV) from the anion framework model, NEB calculated $E_m$ for Li ion migration by the *Tet-Oct-Tet* or *Oct-Tet-Oct* pathways, and $E_m$ from *ab*-initio molecular dynamics (AIMD) simulations for some real lithium compounds with *fcc* anion frameworks.

| Compounds | $E_m$ (Model predicted) | $E_m$ (NEB calculations) | $E_m$ (AIMD simulations) | Comments |
|---|---|---|---|---|
| $LiCoO_2$ | 0.82 | 0.67[3], 0.73[4], 0.83[5] | | |
| $Li_2MnO_3$ | 0.73 | 0.56, 0.72[6] | | NEB calculations by the single-vacancy mechanism; 0.56 for intralayer migration, 0.72 for interlayer migration |
| $Li_4Ti_5O_{12}$ | 0.43 | 0.30, 0.33, 0.36, 0.46, 0.48[7] | | |
| $LiMn_2O_4$ | 0.46 | 0.40, 0.58[8] | | |
| $\gamma$-$Li_3PO_4$ | 0.40 | 0.17, 0.21, 0.23, 0.35[9]; 0.36, 0.45, 0.56, 0.63, 0.69[9] | 0.56[10] | 0.17, 0.21, 0.23, 0.35 by the interstitial mechanism, 0.36, 0.45, 0.56, 0.63, 0.69 by the vacancy mechanism |
| $P\bar{3}m1$-$LiTiS_2$ | 0.73 | 0.70[11], 0.75[12] | | |
| $R\bar{3}m$-$LiTiS_2$ | 0.66 | 0.75[12] | | |
| $\beta$-$Li_3PS_4$ | 0.32 | 0.26[13], 0.30[14] | 0.29[15], 0.31[15] | |
| $Li_3AsS_4$ | 0.27 | 0.22[16], 0.30[17], 0.40[17] | | 0.30 by the interstitial mechanism, 0.40 by the vacancy mechanism |
| $Li_2S$ | 0.50 | 0.47[18], 0.48[19] | | 0.47 by the interstitial mechanism |